\documentclass[referee]{raa}            

\usepackage{graphicx,times}             
\usepackage{natbib}
\usepackage{amssymb,amsmath}
\bibpunct{(}{)}{;}{a}{}{,}
\newcommand{\ba}{\begin{eqnarray}}
\newcommand{\ea}{\end{eqnarray}} 

\usepackage[pagebackref=true]{hyperref}

\begin{document}

  \title{The large scale structure probes of dark energy}

   \volnopage{Vol.0 (202x) No.0, 000--000}      
   \setcounter{page}{1}          

   \author{Pengjie Zhang 
      \inst{1,2,3*}\footnotetext{$*$Corresponding Author.}
      \and Shang Li \inst{1,3}
   }

   \institute{Department of Astronomy, School of Physics and Astronomy, Shanghai Jiao Tong University, Shanghai, 200240, P.R. China; {\it zhangpj@sjtu.edu.cn}\\
        \and
            Division of Astronomy and Astrophysics, Tsung-Dao Lee Institute, Shanghai Jiao Tong University, Shanghai, 200240, P.R. China\\
        \and
             State Key laboratory of dark matter physics, Shanghai Jiao Tong University, Shanghai, 200240, P.R. China\\
\vs\no
   {\small Received 202x month day; accepted 202x month day}}

\abstract{We present a brief review on the large scale structure (LSS)
  probes of dark energy. We cover probes that directly constrain dark
  energy such as baryon acoustic oscillation, redshift space
  distortion, weak lensing and cluster number count. We also review
  auxiliary probes that mitigate systematics in dark energy
  constraints, such as the SZ effect to constrain baryonic effect and
  broadband galaxy clustering to calibrate photometric redshift. We
  demonstrate the synergy between these probes in delivering dark
  energy 
  constraint of both high precision and high accuracy. 
}

   \authorrunning{Zhang and Li}            
   \titlerunning{LSS probes of dark energy}  

   \maketitle

%
%

\section{Introduction}           
\label{sec:intro}

The discovery of the late-time cosmic acceleration and dark energy at the end
of the twentieth century fundamentally altered our understanding of
cosmology \citep{1984ApJ...284..439P,
  1990Natur.348..705E,1995GReGr..27.1137K, 1998AJ....116.1009R, 1999ApJ...517..565P}. The
most straightforward interpretation of the observed cosmic
acceleration within general relativity (GR) is
the cosmological constant \(\Lambda\), with an equation of state \(w \equiv p/\rho =
-1\). The existence of a $\Lambda$-like dark energy component,
dominating the matter-energy budget of the late-time universe,  has
been robustly confirmed by independent observations, including type Ia supernovae
\citep{1998AJ....116.1009R, 1999ApJ...517..565P,
  2018ApJ...859..101S,2022ApJ...938..113S}, cosmic microwave
background \citep{2011PhRvL.107b1302S,
  2014A&A...571A..16P,2020A&A...641A...1P} and baryon acoustic
oscillation \citep{2005ApJ...633..560E, 2021MNRAS.504...33Z,
  2021PhRvD.103h3533A}.  As a consequence, the $\Lambda$CDM cosmology, with $\Lambda$
being a key ingredient, has become the standard model of modern physical cosmology. 

However, recent data, in particular the DESI BAO measurements
\citep{2025JCAP...02..021A,2025PhRvD.112h3515A,2025PhRvD.112h3511L,2025NatAs...9.1879G} in combination with CMB and SNe Ia data, reported $3$-$4\sigma$ evidence for a dynamical dark energy of quintom-like behavior \citep{2005PhLB..607...35F}. These observational
evidences, together with the longstanding fine-tuning problem and the
coincidence problem from the theory side, are challenging the
$\Lambda$CDM paradigm. A dynamical dark energy not only affects the
expansion rate of the universe. It also impacts the large scale
structure (LSS) of the universe, either indirectly through the expansion
rate, or directly by inducing new sources of
inhomogeneities. Furthermore, it may be a hint for the failure of GR
and hence the need for modified gravity  at cosmological
scales. Distinguishing these possibilities requires not only probes of
the overall expansion rate of the universe such as SNe Ia, but also
probes of the structure growth rate \citep{2008PhRvD..78f3503J,2013PhR...530...87W, 2019ARA&A..57..335F,
  2019LRR....22....1I}.  For this purpose,  LSS probes of dark energy,  capable of  measuring both the expansion
rate and structure growth rate,  are indispensable.

Our understanding of LSS probes of dark energy has been deepening  over the last
decade. Theoretical modeling and observational measurements are
improving. Tensions rise and fall  \citep{2025PDU....4901965D}, while systematics  in both theory and measurement are being scrutinized.  New probes, new applications and
new statistics are emerging. These developments motivate us to write a
brief review, structured as follows. Section~\ref{sec:theory} provides a concise theoretical background on dark energy, including its classification and parameterization. Section~\ref{sec:direct} examines direct LSS probes in detail, while Section~\ref{sec:auxiliary} discusses auxiliary probes. Section~\ref{sec:conclusion} presents a brief summary and discussion.  Table \ref{tab:probes} summarizes the probes mentioned in this paper and Fig. \ref{fig:sensitivity} shows the sensitivity of these LSS probes to dark energy properties and synergies between these probes. 

\section{Dark energy}
\label{sec:theory}
The underlying mechanism of the observed cosmic acceleration  can be broadly divided into two categories: dark energy (DE) models that introduce a new cosmic fluid with negative pressure, and modified gravity (MG) models that alter the gravitational interaction itself. A distinction between the two can be drawn based on the  equivalence principles: DE models respect the weak and strong equivalence principle, whereas MG models typically violate the strong equivalence principle and/or even the weak equivalence principle \citep{2016ARNPS..66...95J}. A comprehensive review of dark energy and modified gravity models, such as quintessence, phantom, quintom, k-essence, $f(R)$ gravity, DGP, and more general models such as the Horndeski theory, can be found in review articles of \citep{2012PhR...513....1C,2015PhR...568....1J,2016ARNPS..66...95J,2019ARA&A..57..335F, 2019LRR....22....1I}. 

Alternatively, we could parameterize their observational consequences at the background and perturbation level independently.   For brevity, hereafter we adopt a flat cosmology and the metric 
\begin{equation}
d\tau^2=(1+2\Psi)dt^2-a^2(t)(1-2\Phi)\delta_{ij}dx^i dx^j\ .
\end{equation}
Throughout this paper we set the speed of light $c=1$. 
Different dark energy/modified gravity models have different impacts on observables of the expansion rate such as $H(z)$, $D_A(z)$, and observables of the structure growth rate such as $\Psi$, $\Phi$, the matter overdensity $\delta_m$ and peculiar velocity ${\bf v}$ \citep{2008PhRvD..78f3503J}. 
\begin{itemize}
\item Parametrization at the background level. Both smooth dark energy and clustered dark energy/modified gravity models need it.  Smooth dark energy has negligible inhomogeneities in its energy-momentum tensor. An example is quintessence, whose sound speed $c_{\rm s}=1$, making its distribution smooth on sub-horizon scales. It can be fully parametrized by  the present day density $\Omega_{\rm DE}$ and the equation of state $w(a)$.   
\ba
\label{eqn:H2}
H^2=H_0^2\left(\Omega_{m}a^{-3}+\Omega_{\rm DE}\exp\left(3\int_a^1 (1+w(a))\frac{da}{a}\right)\right)\ \ .
\ea
Note that $\Omega_m+\Omega_{\rm DE}=1$ since we adopt a flat universe. 
In the literature, $w(a)$ is often parameterized as $w(a)=w_0+w_a(1-a)$ with two parameters $w_0$ and $w_a$ \citep{2001IJMPD..10..213C,2003PhRvL..90i1301L}. Although motivated by smooth dark energy, Eq.  \ref{eqn:H2} is also used to parameterize the impact of clustered dark energy and modified gravity on the background expansion. 
 \item Parametrization at the perturbation level. For smooth dark energy, its impact on the structure formation is fully determined by its impact on the expansion rate.   For example, in the $\gamma$ parameterization \citep{1980lssu.book.....P}, $f\equiv d\ln \delta_m/d\ln a=\Omega^\gamma_m(z)$, and $\gamma\simeq 0.55+0.05(1+w(z=1))$ \citep{2005PhRvD..72d3529L}.  However, for clustered dark energy/modified gravity, we need extra parameterization.   When the dark energy sound speed $c_{\rm s}\ll 1$, it can cluster on sub-horizon scales and therefore imprints in the perturbation observables $\Psi$, $\Phi$, $\delta_m$ and ${\bf v}$. Furthermore, it may have non-negligible anisotropic stress $\sigma$, which is a direct gravitational source of metric perturbation and drives the gravitational slip parameter $\eta\equiv \Phi/\Psi\neq 1$. Correspondingly, there are two degrees of freedom to modify GR at the linear scalar perturbation level \citep{2007PhRvL..99n1302Z}. One is to modify the $\Psi$-$\delta_m$ relation (the cosmological Poisson equation) with an effective gravitational constant $G_{\rm matter}$ (or equivalently $\mu(k,z)\equiv G_{\rm matter}/G$).  The other is to modify  the lensing/Weyl potential $\Psi+\Phi$-$\delta_m$ relation with $G_{\rm light}=G\Sigma(k,z)$.  On sub-horizon scales where the quasi-static limit applies, 
 \ba
 \nabla^2\Psi=4\pi G\mu \rho_m a^2\delta_m\ ,\ \nabla^2(\Psi+\Phi)=8\pi G\Sigma\rho_m a^2\delta_m\ .
 \ea
 Note that the gauge invariant $\Delta_m$ instead of $\delta_m$ is often used in the above relations. However, since the two relations are already approximations on sub-horizon scales, the distinction between $\Delta_m$ and $\delta_m$ is unnecessary.  Correspondingly, $\eta(k,z)=2\Sigma/\mu-1$.
Since both clustered dark and modified gravity are modifications to Einstein's field equation and we can move them from one side of the equation to the other side, they are degenerate in their observational consequences \citep{2007PhRvL..98l1301K}.   
\end{itemize}
In the above parameterization of dark energy, we need $\Omega_{\rm DE}$, $w(a)$, $c_{\rm s}$, $\sigma$, or equivalently, $\Omega_{\rm DE}$, $w(a)$, $G_{\rm matter}/G$ ($\mu$) and $G_{\rm light}/G$ ($\Sigma$).  Note that for a flat universe, $\Omega_{\rm DE}=1-\Omega_m$, not an independent parameter. When we detect $w\neq -1$, we obtain observational evidence of dark energy. When we further find $\mu\neq 1$ or $\Sigma\neq 1$, we can conclude that dark energy must be clustered or GR must be modified at scales of interest. 
Dark energy and modified gravity can also be parameterized in a unified manner through the Effective Field Theory of Dark Energy  \citep{2013JCAP...02..032G,2014PhRvD..89j3530H}, which can be translated into $\mu$ and $\Sigma$ \citep{2014JCAP...07..050B}.  

\begin{figure}
   \centering
   \includegraphics[width=1.0\textwidth]{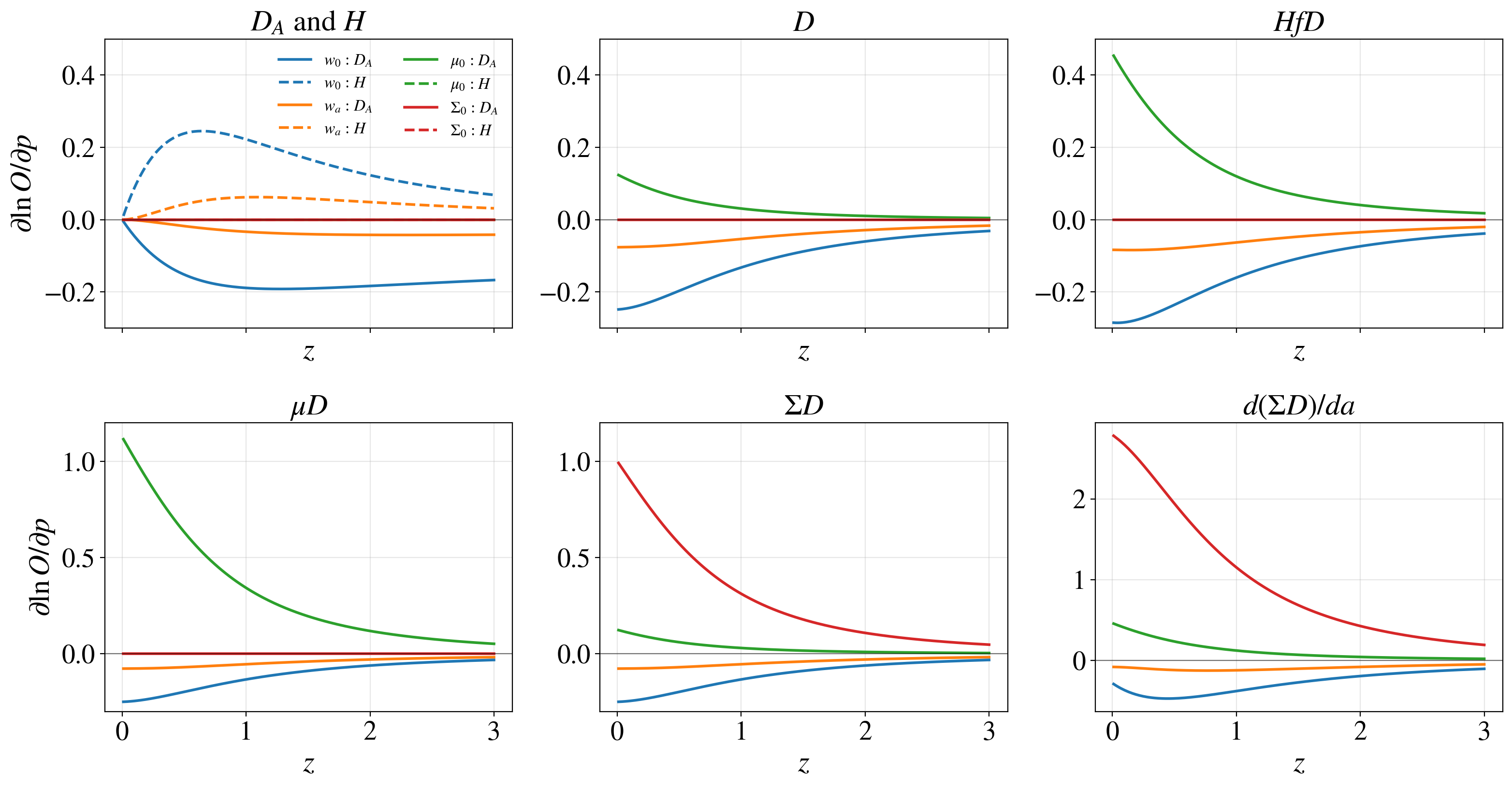}
   \caption{Sensitivity of different observables to dark energy parameters. Here the observables $O$ are chosen as $D_A$, $H$, the linear density growth factor $D$, peculiar velocity $\propto HfD$, weak lensing $\propto \Sigma D$, relativistic effect proportional to gravitational redshift $\Psi\propto \mu D$ and the decay rate of gravitational potential $\propto d(\Sigma D)/da$. The dark energy parameters $p$ are chosen as $w_0$, $w_a$, $\mu_0$ and $\Sigma_0$.  The amplitude $\partial \ln O/\partial p$ quantifies the sensitivity of $O$ to $p$, while the shapes (including the signs) quantify the synergy between different observables. }
   \label{fig:sensitivity}
   \end{figure}

Hereafter we will adopt the $\mu$-$\Sigma$  parameterization, which is more convenient to implement than the $c_{\rm s}$-$\sigma$ parameterization. The background/perturbation observables and parameters to describe them can be summarized into the following equation, 
\ba
\label{eqn:observable}
\underbrace{D_A,H,\cdots}_{\rm background\  observables}\ +\ \underbrace{\Psi,\Phi,\delta_m,v_{\parallel},\cdots}_{\rm perturbation\  observables} \leftrightarrow \underbrace{\Omega_{\rm DE},w_0,w_a,\cdots }_{\rm background\  parameters}\ +\ \underbrace{\mu_0,\Sigma_0,\cdots}_{\rm perturbation\  parameters}\ .
\ea
when neglecting the possible $k$ dependence, the time/redshift dependences of $\mu$ and $\Sigma$ are often parameterized as $\mu(a)=1+\mu_0 \Omega_{\rm DE}(a)/\Omega_{\rm DE}(a=1)$ and $\Sigma(a)=1+\Sigma_0 \Omega_{\rm DE}(a)/\Omega_{\rm DE}(a=1)$.   Fig. \ref{fig:sensitivity} shows the sensitivity of the observables in Eq. \ref{eqn:observable} to dark energy parameters ($w_0$, $w_a$, $\mu_0$ and $\Sigma_0$).  

Table \ref{tab:probes} presents an incomplete summary of LSS probes of dark energy.   Joint analysis of multiple probes is essential for robust dark energy constraints, since there are layers of degeneracy to break. For example, with only $D_A(z)$ or $H(z)$, there is a degeneracy between DE and curvature $\Omega_K$, if we allow $w(a)$ to be a free function of $a$. The combination of $D_A$ and $H$ allows for DE-agnostic determination of $\Omega_K$ solely based on the FRW metric.  Expansion rate measurements alone can distinguish specific DE models from specific MG models. However, more general DE and MG models can generate identical expansion rate history. Structure growth measurements can furthermore break this DE-MG degeneracy. However, at the field equation level, there is no clear distinction between DE and MG. So eventually we aim to measure at least three of the four perturbation variables ($\Psi$, $\Phi$, $\delta_m$ and ${\bf v}$) in order to obtain $G_{\rm matter}$ and $G_{\rm light}$  \citep{2008PhRvD..78f3503J}.  This is observationally challenging, however possible with comprehensive LSS probes  and deeper understanding of them (e.g., \citet{2025arXiv251011022Z}). 

A further signature of clustered dark energy and modified gravity is its environment-dependent behavior. This smoking gun evidence can in principle be tested over a variety of scales ranging from stars to cosmological scales \citep{2021RvMP...93a5003B}. A prerequisite to implement such tests is robust determination of environment (e.g., density or gravitational potential). Future cosmological surveys will map galaxies, stars and diffuse matter distribution to unprecedented width, depth and precision, and therefore significantly enhance this kind of DE/MG test. For brevity, we will not discuss further on this aspect. 

\begin{table}
\begin{center}
\caption[]{LSS probes of dark energy. Information about dark energy is encoded either in the expansion rate of the universe ($H(z)$ and $D_A(z)$), or the large scale structure of the universe (the metric perturbation $\Psi$ and $\Phi$, the matter overdensity $\delta_m$ and velocity ${\bf v}$). If a probe directly contains information about the above properties, it is classified as a direct probe of dark energy.  Auxiliary probes do not contain such direct information, instead they contain  information of   mitigating systematics or breaking parameter degeneracies.  }\label{tab:probes}
\begin{tabular}{lll}
  \hline\noalign{\smallskip}
{\bf Direct probes} & Information contained &  References/remarks                    \\
  \hline\noalign{\smallskip}
Features in galaxy clustering: & &\\
Baryon acoustic oscillation &  $D_A(z)$ \& $H(z)$ &  e.g., \citet{2025JCAP...02..021A}\\  
Redshift space distortion &  $v_\parallel$ & e.g., \citet{2025JCAP...09..008A}\\

Turnover scale & $D_A(z)$ \&  $H(z)$  & e.g., \citet{2026JCAP...04..004Z}\\  
Relativistic dipole & $\Psi$, $v_\parallel$, $\dot{v}_\parallel$ & Emerging probe: e.g., \citet{2012PhRvD..85b3504J,2024arXiv240305398M}\\
\hline
Cosmic shear   & $\Psi+\Phi$ &   e.g., \citet{2026arXiv260210065D}\\
CMB lensing    &  $\Psi+\Phi$   &     e.g., \citet{2011PhRvL.107b1301D,2026PhRvL.136b1001Q}\\
Magnification bias & $\Psi+\Phi$ & e.g., \citet{2005ApJ...633..589S,2024ApJ...973..102X,2025PhRvD.112d3522Q} \\
Lensing magnification & $\Psi+\Phi$ & Emerging probe using SNe Ia or bright sirens\\

\hline
Lensing ratio  & $D_A$ &  \citet{2003PhRvL..91n1302J,2005ApJ...635..806Z,2022PhRvD.105h3529S} \\
N$\times$2pt & $\Psi+\Phi$ & e.g., \citet{2025PhRvD.111j3540S,2026arXiv260114559D}\\
\hline
Integrated Sachs-Wolfe effect & $\dot\Psi+\dot{\Phi}$ & e.g., \citet{2022ApJ...938...72D}\\
Galaxy cluster number count & $\delta_m$ &  e.g., \citet{Planck2015cluster,eROSITA25} \\
Peculiar velocity & $v_{\parallel}$ & e.g., \citet{2023ApJ...944...94T,2024MNRAS.528.4922S,Qin26}\\
21cm intensity mapping & $D_A$, $H$, $v_{\parallel}$ & e.g., \citet{2008PhRvL.100i1303C,2010Natur.466..463C,2025arXiv251119620C}\\
Cluster gravitational redshift & $\Psi$  & Emerging probe: e.g., \citet{2011Natur.477..567W,2024arXiv240305398M} \\
Dark siren clustering & $D_A(D_L)$ \& $H(D_L)$ & Emerging probe: luminosity-distance space LSS \\ 
\hline
{\bf Auxiliary probes} & & \\
\hline
Thermal SZ effect &  ionized diffuse gas: $n_eT_e$ & Calibrating baryonic effects (e.g., \citet{2023ApJ...953..188C})\\
Kinetic SZ effect & ionized diffuse gas: $n_ev_{\parallel}$ & Calibrating baryonic effects (e.g., \citet{2025PhRvD.112h3509H})\\
  & & Testing the Cosmological Principle (e.g., \citet{2011PhRvL.107d1301Z})\\
FRB dispersion measure & ionized diffuse gas: $n_e$   & Calibrating baryonic effects: emerging probe\\
Galaxy and 21cm surveys &  other baryon components & Calibrating baryonic effects\\
Galaxy intrinsic alignment & $\propto \Psi$ & Cosmic shear systematics and LSS signal \\
Broadband galaxy clustering & $P_g(k)$, $C_g(l)$, etc.  & Photo-z calibration \\
\hline

  \noalign{\smallskip}\hline
\end{tabular}
\end{center}
\end{table} 

\section{Direct LSS probes of dark energy}
\label{sec:direct}
Among the variety of LSS probes of dark energy, some of them directly probes $D_A$, $H$, $\Psi$, $\Phi$, $\delta_m$ and ${\bf v}$ and therefore directly contribute to dark energy constraints. We classify them as direct LSS probes of dark energy. Other probes do not contribute directly to the dark energy constraints. However, when analyzing jointly, they are able to either break parameter degeneracies, mitigate systematics or test fundamental principles of cosmology. We classify them as auxiliary probes. We also caution that the classification is not strict, since some probes can both contribute directly and indirectly to dark energy constraint, such as galaxy clustering, the SZ effect and galaxy intrinsic alignment. 

\subsection{Features in galaxy clustering}
The analysis of galaxy clustering in the context of cosmology is used to focus on features in the galaxy clustering, such as the BAO bump, Alcock-Paczynski and RSD anisotropies in the correlation function. Recently, full-shape analysis has become the standard analysis method \citep{2025JCAP...09..008A} and simulation-based inference and field-level inference  are drawing more attentions. For brevity, we focus on features in the galaxy clustering. 
\begin{itemize}
\item Baryon acoustic oscillation (BAO). This is the relic of primordial density fluctuations at the drag epoch recorded in the galaxy clustering, manifesting as a bump in the galaxy correlation function at the drag epoch sound horizon $r_{\rm d}\sim 150 $ Mpc. It serves as a standard ruler to measure the angular diameter distance $D_A/r_{\rm d}$ and the Hubble distance $1/(Hr_{\rm d})$, both in unit of $r_{\rm d}$. It was first detected by SDSS \citep{2005ApJ...633..560E} and 2DF \citep{2005MNRAS.362..505C}. The  complete SDSS survey improved the total distance measurement accuracy to $\sim 0.7\%$ \citep{2021PhRvD.103h3533A}, while DESI further improved to $0.3\%$ \citep{2025PhRvD.112h3515A}. Furthermore, combining all tracers, BAO measurements of DESI span a redshift range of $z<2.33$.  
\item Redshift space distortion (RSD). This is the peculiar velocity induced anisotropies in the galaxy clustering. Galaxy peculiar velocity alters the observed galaxy redshift $z$ to $z+v_{\parallel}$. So the observed redshift space galaxy position is ${\bf s}={\bf x}+v_\parallel \hat{x}/H$. The coherent part of the velocity (velocity gradient) enhances the clustering (the Kaiser effect), while the random part suppresses the clustering (the Finger of God effect). However, accurate modeling of RSD is challenging, suffering three layers of nonlinearity in the real space-redshift space mapping, matter density-velocity relation and matter-galaxy relation. By the continuity equation and vanishing velocity bias \citep{2018ApJ...861...58C}, $v\propto f\sigma_8(z)$. Therefore RSD measurement is often compressed into a single parameter $f\sigma_8(z)$. Here $f\equiv d\ln D/d\ln a$ and $D$ is the linear density growth factor. RSD measurements have progressed from early pioneering constraints to precise, multi-tracer cosmological probes. For example, the complete SDSS surveys achieved a combined precision of  $4.8\%$ over the redshift interval $z <1.5$ \citep{2021PhRvD.103h3533A}. Furthermore, a multi-tracer analysis  was for the first time realized using LRG and ELG galaxy samples \citep{2021MNRAS.504...33Z}, demonstrating the potential of cosmic variance cancellation. DESI, on the other hand,  has achieved a combined precision of $4.7\%$ just using the first year data of $4.7$ million galaxy and quasar redshifts over $0<z<2.1$ \citep{2025JCAP...09..008A}. 
\item Turnover scale. Since density perturbations grow differently on the sub-horizon and super-horizon scales and in the radiation/matter dominance epoch, the horizon $r_{\rm eq}$ of the matter-radiation equality epoch imprints a turnover scale in  the matter power spectrum.  This feature at $r_{\rm eq}\sim 680$ Mpc is best measured with surveys of large sky coverage and wide redshift depth, such as DESI \citep{2026JCAP...04..004Z}. Although its precision is lower, it serves as an independent standard ruler. 
\end{itemize}
There are also other features in the galaxy distribution useful for cosmology such as voids. Voids are low-density counterparts of galaxy clusters. Analogous to cluster mass function, one can probe dark energy through the void size function. Furthermore, voids are alternative to galaxies to perform LSS analysis such as void clustering and void lensing. For example, it can be used to enhance BAO and structure growth measurement \citep{2022MNRAS.511.5492Z,2022MNRAS.513..186A}. 

 The complete DESI survey and future stage V spectroscopic redshift surveys \citep{2024arXiv240305398M,2024arXiv241107970Z,2025arXiv250307923B} will measure BAO and RSD with higher precision while extending to  higher redshifts. Stage V surveys and beyond have the capability to measure the relativistic dipole for the first time \citep{2023MNRAS.525.4611B,2024arXiv240305398M}. Gravitational redshift and peculiar velocity (not its gradient) modify the observed galaxy  overdensity with additive terms proportional to $\nabla_\parallel \Psi$, $v_\parallel $ and $\dot{v}_\parallel$. These terms are suppressed  by $H/k$ with respect to density and RSD, and are non-negligible only at very large scales. A major reason that they can in principle be measured is that they generate a dipole in the correlation function. There are other features in the galaxy clustering, such as the magnification bias, which will be discussed in the weak lensing section. 

A full exploitation of these features requires a large volume galaxy spectroscopic surveys. However, imaging surveys with sufficiently robust photometric redshift measurements are also capable of extracting some of these features, in particular BAO. For example, DES year 6 measured  the angular BAO and obtained a $2\%$ measurement of $D_A$  \citep{2024PhRvD.110f3515A} at an effective redshift $z_{\rm eff}=0.85$. Furthermore, BAO can be measured from the cross-correlation between spectroscopic and photometric samples (e.g., refer to \citet{2024MNRAS.527.3728D} for a CSST forecast).

\subsection{Weak gravitational lensing}
Weak gravitational lensing probes the light deflection, so it is in principle not only able to probe the scalar metric perturbations, but also the vector and tensor metric perturbations. For the scalar perturbations, the measured signal is the lensing convergence (or equivalently the shear) $\kappa=\int W_L(z_{\rm L},z_{\rm s})\nabla^2(\Psi+\Phi) dz$. Here $W_L$ is the lensing kernel, depending both on the source redshift $z_{\rm s}$ and the lens redshift $z_{\rm L}$.  Current surveys have measured weak lensing  to high significance, with independent lensing measurement methods. 
\begin{itemize}
\item Cosmic shear. This is the lensing induced coherent distortion of galaxy images. Given a typical signal of $1\%$ at typical source redshift $z_{\rm s}=1$ and typical shape noise of $30\%$, it requires at least $\sim 10^3$ galaxy images to achieve $S/N\sim 1$. Cosmic shear in a blank sky was first detected in the year 2000 with $\sim 10^5$ galaxies. Current stage III surveys such as DES, HSC and KiDS have measured $\sim 10^8$ galaxy images to $z_{\rm s}>1$. HSC year 3 data has $S/N\sim 27$ \citep{2023PhRvD.108l3518L}, while DES year 6 data reached S/N$\sim 83$ \citep{2026arXiv260114559D}. Stage IV surveys such as CSST, Euclid and LSST will improve the S/N to $\gg 10^2$ (e.g., \citet{2024MNRAS.527.5206Y}), while extending to higher source redshifts. 
\item CMB lensing. This is the lensing induced  secondary anisotropy in the CMB temperature and polarization fields. It was first detected at $4\sigma$ by ACT \citep{2011PhRvL.107b1301D}, which enabled the confirmation of dark energy by CMB alone \citep{2011PhRvL.107b1302S}. Planck has dramatically increased the S/N to $42$ (Planck PR4, \citet{2020A&A...641A...1P}), followed by ACT DR6 with S/N of $43$ \citep{2024ApJ...962..112Q} and SPT-3G with S/N of $38$ \citep{2025PhRvD.111h3534G}. The three measurements are almost independent and the combined S/N is $61$ \citep{2026PhRvL.136b1001Q}.  CMB lensing measurement is under the transition from temperature based reconstruction to polarization-based reconstruction. Ongoing Simons Observatory \citep{2019JCAP...02..056A} and more advanced future experiments will eventually push the lensing reconstruction to the cosmic variance limit at $L\la 10^3$ and improve the S/N to $\gg 10^2$ . 
\end{itemize}

Besides the two sophisticated methods, there are emerging probes. (1) One is the Magnification bias. Weak lensing also modifies the galaxy number overdensity through both the flux magnification and area amplification,  $\delta_g\rightarrow \delta_g+2(\alpha-1)\kappa$. A unique feature of magnification bias is that it induces a spatial correlation between two galaxy(quasar) samples widely separated in redshift, and has enabled its first detection \citep{2005ApJ...633..589S}. Another unique feature is the $2(\alpha-1)$ prefactor, which depends on the galaxy selection criteria and is an observable for a flux-limited sample. Utilizing this characteristic dependence, \citet{2025PhRvD.112d3522Q} has reconstructed a lensing map by counting DES galaxies.  (2) Another is the flux magnification. This can be measured from the average flux/magnitude (e.g., \citet{2024ApJ...973..102X}), magnitude fluctuations of SNe Ia, or fluctuations in the luminosity-distance of bright standard sirens. 

\subsection{Galaxy clustering $\times$ weak lensing}
A major disadvantage of CMB lensing, comparing to tomographic cosmic shear, is the lack of redshift evolution due to the single source redshift at the epoch of recombination. This can be overcome by cross-correlating with galaxies/quasars.  Together with the galaxy auto-correlation, this recovers $\sigma_8(z)$ at the galaxy/quasar redshifts (e.g., \citet{2025PhRvD.111j3540S,2025JCAP...11..077M,2026ApJ..1001..121Q}). Note that in the context of clustered dark energy and modified gravity, these measurements constrain the combination $\Sigma\sigma_8(z)$ instead. Even in the case of cosmic shear, cross-correlating with galaxies (e.g., the widely used 3$\times$2pt analysis in cosmic shear \citep{2026arXiv260210065D}) breaks the degeneracy between galaxy bias and matter density fluctuation, and is therefore highly valuable.  The 3$\times$2pt (shear auto-correlation, shear-galaxy cross-correlation and galaxy auto-correlation) can be extended to N$\times$2pt by including other cross-correlations such as galaxy-CMB lensing cross-correlation and cosmic shear-CMB lensing cross-correlation. 

The above analysis requires the full-shape analysis of summary statistics. Alternatively, by fixing the lens redshift while varying the source redshift, one can isolate a pure geometric measure $R=W_L(z_{\rm L},z_{\rm s}^A)/W_L(z_{\rm L},z_{\rm s}^B)$ \citep{2003PhRvL..91n1302J,2005ApJ...635..806Z}. Since it does not rely on modeling of correlation function, it is applicable to small scale regardless of nonlinearity and baryonic effects \citep{2022PhRvD.105h3529S}. Including CMB lensing at the highest source redshift $\sim 1100$, its constraining power can be further enhanced. 

\subsection{Galaxy cluster number count}
\label{subsec:cluster}
As rare density peaks of the underlying matter field, the abundance of galaxy clusters is highly sensitive to $\delta_m(z)$ and therefore $\sigma_8(z)$. The first task in cluster cosmology is to construct cluster catalogs from optical surveys, SZ surveys, X-ray surveys or weak lensing surveys.  
\begin{itemize}
\item SZ selected clusters. Being the widest SZ surveys, Planck 2015 selected 1653 clusters  down to a signal-to-noise of 4 over $84\%$ of the sky \citep{2016A&A...594A..27P}, and the cosmological analysis used 439 clusters  down to S/N of $6$ \citep{Planck2015cluster}. Being the deepest SZ surveys, SPT (SPT-SZ, SPTpol ECS and SPTpol 500d) constructed a catalog of 1,005 confirmed clusters with S/N$>5$  in the redshift range $0.25$-$1.78$  over a total sky area of 5,200 deg$^2$ \citep{2024PhRvD.110h3510B}. ACT DR6 detected 10040 clusters with S/N$>4$ covering $16293$  deg$^2$ \citep{2026OJAp....955863A}.  The catalog includes 1180 clusters at $z>1$ and 124 clusters at $z>1.5$. Note that the mass threshold of SZ selected clusters is approximately redshift independent. For example, the $90\%$ mass completeness limit of the ACT DR6 catalog is  approximately $M_{\rm 500c}=5\times  10^{14} M_\odot$ at $z>2$. As a comparison, this number is $1.5\times 10^{14}M_\odot$ for SPT. 
\item X-ray selected clusters. X-ray surveys such as ROSAT all-sky survey can produce a large amount of X-ray selected clusters (e.g. \citet{2019MNRAS.488..739K}) Using eROSITA surveyed in its first six months of operation, a catalog of 5259 securely detected and confirmed clusters  was constructed \citep{2024A&A...685A.106B}. It locates in the 12791 deg$^2$ common footprint with the DESI Legacy Survey DR10 and has a higher purity level ($\sim 95\%$), making it suitable for subsequent cosmological study \citep{2024A&A...689A.298G,eROSITA25}. 
\item Optically selected clusters.  DES has constructed a redMaPPer cluster catalog with about 16000 clusters to $z<0.65$, using year 3 data \citep{2025PhRvD.112h3535A}). Based on DESI imaging surveys DR8, \citet{2021ApJ...909..143Y} constructed a halo-based group catalog over $18,000$ deg$^2$ sky to $z<1$. About 500,000 of these groups have at least 10 members. Among this catalog, clusters  have mass estimation of better than 0.2 dex uncertainty.  
\item Weak lensing selected clusters. Given the finite angular size of galaxy clusters and the limited number density of source galaxies behind them, the construction of a large shear selected clusters requires a deep lensing survey.  The Subaru Hyper Suprime-Cam survey, being the deepest ($\sim 20$ galaxies per arcmin$^{2}$) among stage III surveys, has identified more than 300  clusters with S/N greater than 5 \citep{2025PJAB..101..129O}.  The number of shear selected clusters will be increased by a factor of $\sim 10^2$ by stage IV surveys. Alternatively, one can use weak lensing peaks  instead of confirmed clusters to enhance cosmological constraints \citep{2010ApJ...719.1408F,2025PJAB..101..129O}.  
\end{itemize}
A major issue challenging precision cluster cosmology is the cluster mass determination (or the mass-observable relation). Weak lensing is a major tool to calibrate mass of clusters selected optically, via SZ or X-ray. A related, but not identical, issue is the determination of cluster selection function.  It in general depends on properties other than the cluster mass and is further complicated by astrophysical contaminations and observational issues \citep{2023MNRAS.522.1601C,2025PJAB..101..129O}.  On the other hand, cluster cosmology is of great potential, since the cluster mass function is highly sensitive to dark energy and its dependence on $w_0$ and $w_a$ is  known to high accuracy (e.g., the CSST cosmological emulator  \citep{2025SCPMA..6809513C}).  

\subsection{Other direct LSS probes of dark energy}
Among many other direct LSS probes of dark energy, here we list a few of them.  (1) The integrated Sachs-Wolfe effect in the CMB temperature anisotropies, which provides a unique probe of $\dot{\Psi}+\dot{\Phi}$. However, this signal is overwhelmed by the primary CMB anisotropies and can only be measured through cross-correlation with LSS. Therefore only until full-sky CMB surveys and wide galaxy surveys became available, its measurement becomes reliable. In combination with CMB lensing, one can isolate the gravitational potential decay rate $DR\propto (\dot{\Psi}+\dot{\Phi})/(\Psi+\Phi)$,  which is highly sensitive to dark energy. $DR$ has been measured combining Planck and DESI imaging surveys and has improved the dark energy constraint of SDSS BAO by $30\%$ \citep{2022ApJ...938...72D}. (2)  The galaxy intrinsic alignment, which is the spatially correlated galaxy ellipticities. This is widely regarded as one of the major systematics in cosmic shear cosmology. However, it also contains valuable information about BAO \citep{2023NatAs...7.1259X} and structure growth \citep{2022PhRvD.106d3523O}.  (3) An emerging probe of LSS is binary stellar mass black hole gravitational wave (GW) events. These dark sirens enable direct  luminosity-distance measurements and therefore depict  the luminosity-distance space LSS \citep{2018arXiv181011915Z,2022JCAP...02..003L}.  This unique LSS is measurable by third generation GW experiments \citep{2026JCAP...03..081A}. Furthermore, bright sirens of binary neutron star systems will enable direct measurement of peculiar velocity, analogous to peculiar velocity by SNe Ia. 

\section{Auxiliary probes}
\label{sec:auxiliary}
So far we focus on the cosmological information encoded in various probes. For accurate determination of dark energy property, we shall also diagnose and mitigate systematics  in modeling and measurement of the corresponding statistics. For example,  the baryonic effect, galaxy intrinsic alignment and photo-z errors can all significantly bias the dark energy constraint from weak lensing.  Another source of systematics in dark energy is uncertainties in fundamental principles, such as the cosmological principle (CP).  A weak CP violation could bias the inference toward a dynamical dark energy even if the true model is $\Lambda$CDM.  The kinetic SZ, with distant electrons observing the universe from different viewing angles, is able to test CP model-independently.  For these purposes, there are a variety of auxiliary probes and some emerging probes are of great potential. For example, the dispersion measure (DM) of fast radio bursts (FRBs) contains information about the distribution of baryons in the intergalactic medium. Current cosmological application is limited by contaminations from the host galaxy DM, while  techniques of LSS help \citep{2014ApJ...780L..33M}.  For brevity, we only focus on the thermal SZ (tSZ) effect, the kinetic/kinematic SZ (kSZ) effect and broadband galaxy clustering (instead of features such as BAO).  

\subsection{The thermal  Sunyaev Zel'dovich (tSZ) effect}
The  SZ effect is a secondary CMB anisotropy caused by energy exchange between CMB photons and free electrons through the inverse-Compton scattering (\citet{2002ARA&A..40..643C} and references therein). When the energy source is the electron thermal energy, it is called the  thermal SZ (tSZ) effect. The resulting $\Delta T/T=f(\nu)y$ and $y=\int (k_BT_e/m_e) d\tau_e$.  Here $d\tau_e=\sigma_T n_edl$ is the Thomson optical depth. $n_e$ is the free electron number density. $T_e$ is the electron temperature.  $f(\nu)$ is the characteristic frequency dependence of tSZ, which is a key to separate tSZ from primary CMB and foregrounds.  The tSZ effect of massive galaxy clusters can reach $\sim 10^2\mu$K and has been detected in the 1980s. This effect has become a major tool of cluster finding, in particular at high redshifts (\S \ref{subsec:cluster}).  

With multiple frequency coverage from 30 to 857 GHz, Planck has revolutionized this field \citep{2016A&A...594A..22P}.  With arcminute angular resolution and higher sensitivity, ACT and SPT are pushing the tSZ measurement to smaller angular scales (e.g. \citet{2024PhRvD.109f3530C,2026ApJ..1002L..22R}). However, robust measurement of the tSZ in a blank sky is still challenging, in particular due to contaminations of cosmic infrared background (CIB). One way to mitigate the CIB contamination and boosting the tSZ detection is the SZ tomography \citep{2001ApJ...549...18Z}, namely cross-correlating with LSS tracers. These cross-correlations have been measured using galaxies, clusters, weak lensing maps and X-ray maps with S/N in the range of $10$-$100$ (e.g., \citet{2023ApJ...953..188C,2025PhRvD.112h3561L,2025PhRvD.112d3525L,2026JCAP...02..065L} and references therein).  The tSZ measurements (auto- and cross-) can then be directly compared with hydro-simulations or to convert into cluster gas fraction to probe baryonic physics. Strong feedback and significant suppression in the matter clustering are supported by these analyses \citep{2023ApJ...953..188C,2026ApJ..1002L..22R}.  For tSZ to robustly constrain baryonic effect, improvement in correcting CIB contamination and modeling of gas pressure are required. 

\subsection{The kinetic/kinematic Sunyaev Zel'dovich (kSZ) effect}
The kSZ effect is caused by the bulk motion of free electrons. The resulting CMB temperature fluctuation is $\Delta T/T=\int v_{e,\parallel}d\tau_e$. Here $v_{e,\parallel}$ is the peculiar velocity along the line of sight, in unit of the speed of light. Unlike tSZ, kSZ shares the same spectral dependence as primary CMB, making its separation from other CMB components more challenging than tSZ. Current CMB data only reports marginal detection of the kSZ power spectrum. On the other hand, its dependence on the direction of peculiar velocity enables cleaner  measurements, either through  the kSZ tomography by cross-correlating with a template galaxy momentum field \citep{2009arXiv0903.2845H,2011MNRAS.413..628S} or through the pairwise kSZ measurement of galaxies and clusters \citep{2012PhRvL.109d1101H,2022MNRAS.510.5916C}. The kSZ measurement had limited S/N, but recently improved to  S/N$>20$ by kSZ tomography analysis of DESI DR2 spectroscopic galaxies and ACT DR6 CMB map \citep{2026arXiv260404867C,2026arXiv260419744Q,2026arXiv260419745H}.  Recent kSZ measurements also support strong feedback (e.g., \citet{2025PhRvD.112h3509H}).   

A unique cosmological application of the kSZ effect is to probe horizon-scale inhomogeneities, namely test the cosmological principle (CP). A strong CP violation would  invalidate the observational evidence of cosmic acceleration. A weak CP violation could bias $\Lambda$ as a dynamical dark energy. Therefore testing CP is of particular importance for dark energy. If CP is violated and horizon-scale inhomogeneities exist, there will exist an extra bulk flow decoupled to the local density field, resulting in a unique kSZ effect dubbed the {\it linear kSZ effect} \citep{2011PhRvL.107d1301Z,2015JCAP...06..046Z}. This effect has ruled out strong CP violation and confirmed the existence of cosmic acceleration \citep{2011PhRvL.107d1301Z,2014A&A...561A..97P}, and put strong constraints on weak CP violation \citep{2022MNRAS.509.1291C,2024arXiv240805264K}. 

\subsection{Broadband galaxy clustering}
Broadband galaxy power spectrum contains cosmological information. However its extraction heavily relies on modeling of galaxy bias. On the other hand, broadband galaxy clustering has been shown to be a powerful tool of photometric-redshift calibration.  One way to implement it is the cross-calibration method by cross-correlating the photometric sample with a spectroscopic redshift sample in overlapping sky regions \citep{2008ApJ...684...88N}. The measured cross-power spectrum is $b_{\rm s}b_{\rm p}f_{\rm p} C_m$.  So the fraction of photo-z galaxies in the redshift range of spec-z sample $f_{\rm p}$ can be inferred up to uncertainties in the galaxy bias $b_{\rm p}$ of the photo-z sample in this redshift range. This method is widely used to calibrate the photo-z distribution of source galaxies in weak lensing surveys such as HSC \citep{2026JCAP...05..073C} and  DES \citep{2025arXiv251023565D}. Another way is the self-calibration method by cross-correlating galaxies of the same photometric sample but of different photo-z bins \citep{2006ApJ...651...14S,2010MNRAS.405..359Z,2017ApJ...848...44Z}.  It has been applied to DESI imaging surveys \citep{2024JCAP...10..025P}. The two methods are complementary to each other. The self-calibration method does not require a spectroscopic galaxy sample, and does not suffer from the issue of cross-calibration associated with an unknown $b_{\rm p}$. However, it will fail to diagnose a specific type of photo-z error in which the photo-z and spec-z mapping is  monotonic mapping. In contrast, the cross-calibration is  sensitive to such type of photo-z error. Nonetheless, both methods rely on robust galaxy clustering measurements and are sensitive to  imaging systematics, which may induce spurious correlation between galaxy samples separated widely in redshift.

%
 %

\section{Discussion and conclusions}
\label{sec:conclusion}
The large scale structure of the universe provides a variety of dark energy probes, revealing the imprint of dark energy in both the expansion history of the universe and the structure growth history. Measurements of some of these probes have already achieved high accuracy and a famous example is BAO and the evidence of dynamical dark energy largely based on the DESI BAO measurement. Measurements of some other probes, such as weak lensing and redshift space distortion,  will achieve not only high precision but high accuracy in the coming years. Furthermore, emerging probes such as relativistic effects in the horizon-scale galaxy clustering will provide  information  of dark energy in unexplored regime.  With ongoing and upcoming stage IV cosmological surveys (Euclid, LSST/Rubin, CSST and WFIRST/Roman) and even more powerful stage V projects, and with more advanced analysis methods such as non-Gaussian summary statistics and field-level inference, information about dark energy with unprecedented precision and diversity will be extracted in the coming decades.

  Meanwhile, existing tensions \citep{2025PDU....4901965D} and known systematics all emphasize the essential role of systematic diagnosis and mitigation. LSS probes have been used for this purpose.  One example is the baryonic effect in the matter clustering, which is a major systematic error in weak lensing cosmology and the resulting dark energy constraint. This effect is hard to predict robustly from first principles, even with state-of-the-art hydro simulations. LSS probes such as the SZ effect, on the other hand, are providing direct observational constraints. However, their potential in self-calibration of systematics in dark energy constraints is far from being fully explored, and significantly more efforts are needed. Eventually we expect that LSS probes will deliver  high precision and high accuracy dark energy constraint  in the coming decade. 

\begin{acknowledgements}
This work was supported by the National Key R\&D Program of China (2023YFA1607800 and 2023YFA1607801).
\end{acknowledgements}


   


\begin{thebibliography}{99}
\bibitem[Abac et al.(2026)]{2026JCAP...03..081A} Abac, A. and 199 colleagues 2026.\ The Science of the Einstein Telescope.\ Journal of Cosmology and Astroparticle Physics 2026. doi:10.1088/1475-7516/2026/03/081

\bibitem[Abbott et al.(2024)]{2024PhRvD.110f3515A} Abbott, T.~M.~C. and 114 colleagues 2024.\ Dark Energy Survey: A 2.1\% measurement of the angular baryonic acoustic oscillation scale at redshift zeff=0.85 from the final dataset.\ Physical Review D 110. doi:10.1103/PhysRevD.110.063515

\bibitem[Abbott et al.(2025)]{2025PhRvD.112h3535A} Abbott, T.~M.~C. and 95 colleagues 2025.\ Dark energy survey year 3 results: Cosmological constraints from cluster abundances, weak lensing, and galaxy clustering.\ Physical Review D 112. doi:10.1103/3dzh-d8f5

\bibitem[Abdul Karim et al.(2025)]{2025PhRvD.112h3515A} Abdul Karim, M. and 185 colleagues 2025.\ DESI DR2 results. II. Measurements of baryon acoustic oscillations and cosmological constraints.\ Physical Review D 112. doi:10.1103/tr6y-kpc6

\bibitem[Ade et al.(2019)]{2019JCAP...02..056A} Ade, P. and 199 colleagues 2019.\ The Simons Observatory: science goals and forecasts.\ Journal of Cosmology and Astroparticle Physics 2019. doi:10.1088/1475-7516/2019/02/056

\bibitem[Adame et al.(2025)]{2025JCAP...02..021A} Adame, A.~G. and 199 colleagues 2025.\ DESI 2024 VI: cosmological constraints from the measurements of baryon acoustic oscillations.\ Journal of Cosmology and Astroparticle Physics 2025. doi:10.1088/1475-7516/2025/02/021

\bibitem[Adame et al.(2025)]{2025JCAP...09..008A} Adame, A.~G. and 198 colleagues 2025.\ DESI 2024 V: Full-Shape galaxy clustering from galaxies and quasars.\ Journal of Cosmology and Astroparticle Physics 2025. doi:10.1088/1475-7516/2025/09/008

\bibitem[Adame et al.(2025)]{2025JCAP...09..008A} Adame, A.~G. and 198 colleagues 2025.\ DESI 2024 V: Full-Shape galaxy clustering from galaxies and quasars.\ Journal of Cosmology and Astroparticle Physics 2025. doi:10.1088/1475-7516/2025/09/008

\bibitem[Aguena et al.(2026)]{2026OJAp....955863A} Aguena, M. and 99 colleagues 2026.\ The Atacama Cosmology Telescope: DR6 Sunyaev-Zel'dovich Selected Galaxy Clusters Catalog.\ The Open Journal of Astrophysics 9, 55863. doi:10.33232/001c.155863

\bibitem[Alam et al.(2021)]{2021PhRvD.103h3533A} Alam, S. and 98 colleagues 2021.\ Completed SDSS-IV extended Baryon Oscillation Spectroscopic Survey: Cosmological implications from two decades of spectroscopic surveys at the Apache Point Observatory.\ Physical Review D 103. doi:10.1103/PhysRevD.103.083533

\bibitem[Artis et al.(2025)]{eROSITA25} Artis, E. and 31 colleagues 2025.\ The SRG/eROSITA All-Sky Survey: Constraints on the structure growth from cluster number counts.\ Astronomy and Astrophysics 696. doi:10.1051/0004-6361/202452584

\bibitem[Aubert et al.(2022)]{2022MNRAS.513..186A} Aubert, M. and 21 colleagues 2022.\ The completed SDSS-IV extended Baryon Oscillation Spectroscopic Survey: growth rate of structure measurement from cosmic voids.\ Monthly Notices of the Royal Astronomical Society 513, 186--203. doi:10.1093/mnras/stac828
  
\bibitem[Baker et al.(2021)]{2021RvMP...93a5003B} Baker, T. and 10 colleagues 2021.\ Novel Probes Project: Tests of gravity on astrophysical scales.\ Reviews of Modern Physics 93. doi:10.1103/RevModPhys.93.015003

\bibitem[Bellini and Sawicki(2014)]{2014JCAP...07..050B} Bellini, E., Sawicki, I.\ 2014.\ Maximal freedom at minimum cost: linear large-scale structure in general modifications of gravity.\ Journal of Cosmology and Astroparticle Physics 2014. doi:10.1088/1475-7516/2014/07/050

\bibitem[Besuner et al.(2025)]{2025arXiv250307923B} Besuner, R. and 109 colleagues 2025.\ The Spectroscopic Stage-5 Experiment.\ arXiv e-prints. doi:10.48550/arXiv.2503.07923

\bibitem[Bocquet et al.(2024)]{2024PhRvD.110h3510B} Bocquet, S. and 197 colleagues 2024.\ SPT clusters with DES and HST weak lensing. II. Cosmological constraints from the abundance of massive halos.\ Physical Review D 110. doi:10.1103/PhysRevD.110.083510

\bibitem[Bonvin et al.(2023)]{2023MNRAS.525.4611B} Bonvin, C., Lepori, F., Schulz, S., Tutusaus, I., Adamek, J., Fosalba, P.\ 2023.\ A case study for measuring the relativistic dipole of a galaxy cross-correlation with the Dark Energy Spectroscopic Instrument.\ Monthly Notices of the Royal Astronomical Society 525, 4611--4627. doi:10.1093/mnras/stad2567

\bibitem[Bulbul et al.(2024)]{2024A&A...685A.106B} Bulbul, E. and 33 colleagues 2024.\ The SRG/eROSITA All-Sky Survey. The first catalog of galaxy clusters and groups in the Western Galactic Hemisphere.\ Astronomy and Astrophysics 685. doi:10.1051/0004-6361/202348264

\bibitem[Camarena et al.(2022)]{2022MNRAS.509.1291C} Camarena, D., Marra, V., Sakr, Z., Clarkson, C.\ 2022.\ The Copernican principle in light of the latest cosmological data.\ Monthly Notices of the Royal Astronomical Society 509, 1291--1302. doi:10.1093/mnras/stab3077

\bibitem[Carlstrom et al.(2002)]{2002ARA&A..40..643C} Carlstrom, J.~E., Holder, G.~P., Reese, E.~D.\ 2002.\ Cosmology with the Sunyaev-Zel'dovich Effect.\ Annual Review of Astronomy and Astrophysics 40, 643--680. doi:10.1146/annurev.astro.40.060401.093803

\bibitem[Chang et al.(2008)]{2008PhRvL.100i1303C} Chang, T.-C., Pen, U.-L., Peterson, J.~B., McDonald, P.\ 2008.\ Baryon Acoustic Oscillation Intensity Mapping of Dark Energy.\ Physical Review Letters 100. doi:10.1103/PhysRevLett.100.091303

\bibitem[Chang et al.(2010)]{2010Natur.466..463C} Chang, T.-C., Pen, U.-L., Bandura, K., Peterson, J.~B.\ 2010.\ An intensity map of hydrogen 21-cm emission at redshift z\raisebox{-0.5ex}\textasciitilde0.8.\ Nature 466, 463--465. doi:10.1038/nature09187

\bibitem[Chaussidon et al.(2026)]{2026arXiv260404867C} Chaussidon, E. and 51 colleagues 2026.\ Measurement of the galaxy-velocity power spectrum of DESI tracers with the kinematic Sunyaev-Zeldovich effect using DESI DR2 and ACT DR6.\ arXiv e-prints. doi:10.48550/arXiv.2604.04867

\bibitem[Chen et al.(2018)]{2018ApJ...861...58C} Chen, J., Zhang, P., Zheng, Y., Yu, Y., Jing, Y.\ 2018.\ Accurate Determination of Halo Velocity Bias in Simulations and Its Cosmological Implications.\ The Astrophysical Journal 861. doi:10.3847/1538-4357/aaca2f
  
\bibitem[Chen et al.(2022)]{2022MNRAS.510.5916C} Chen, Z., Zhang, P., Yang, X., Zheng, Y.\ 2022.\ Detection of pairwise kSZ effect with DESI galaxy clusters and Planck.\ Monthly Notices of the Royal Astronomical Society 510, 5916--5928. doi:10.1093/mnras/stab3604

\bibitem[Chen et al.(2023)]{2023ApJ...953..188C} Chen, Z., Zhang, P., Yang, X.\ 2023.\ Thermal Energy Census with the Sunyaev-Zel'dovich Effect of DESI Galaxy Clusters/Groups and Its Implication on the Weak-lensing Power Spectrum.\ The Astrophysical Journal 953. doi:10.3847/1538-4357/ace1e2

\bibitem[Chen and Yu(2025)]{2025SCPMA..6809513C} Chen, Z., Yu, Y.\ 2025.\ CSST cosmological emulator II: Generalized accurate halo mass function emulation.\ Science China Physics, Mechanics, and Astronomy 68. doi:10.1007/s11433-025-2764-x

\bibitem[Chevallier and Polarski(2001)]{2001IJMPD..10..213C} Chevallier, M., Polarski, D.\ 2001.\ Accelerating Universes with Scaling Dark Matter.\ International Journal of Modern Physics D 10, 213?223. doi:10.1142/S0218271801000822

\bibitem[CHIME Collaboration et al.(2025)]{2025arXiv251119620C} CHIME Collaboration and 38 colleagues 2025.\ Detection of the Cosmological 21 cm Signal in Auto-correlation at z \raisebox{-0.5ex}\textasciitilde 1 with the Canadian Hydrogen Intensity Mapping Experiment.\ arXiv e-prints. doi:10.48550/arXiv.2511.19620

\bibitem[Chiu et al.(2023)]{2023MNRAS.522.1601C} Chiu, I.-N., Klein, M., Mohr, J., Bocquet, S.\ 2023.\ Cosmological constraints from galaxy clusters and groups in the eROSITA final equatorial depth survey.\ Monthly Notices of the Royal Astronomical Society 522, 1601--1642. doi:10.1093/mnras/stad957

\bibitem[Clifton et al.(2012)]{2012PhR...513....1C} Clifton, T., Ferreira, P.~G., Padilla, A., Skordis, C.\ 2012.\ Modified gravity and cosmology.\ Physics Reports 513, 1--189. doi:10.1016/j.physrep.2012.01.001

\bibitem[Choppin de Janvry et al.(2026)]{2026JCAP...05..073C} Choppin de Janvry, J. and 55 colleagues 2026.\ Full calibration of the tomographic redshift distribution from the HSC PDR3 Shape Catalog with DESI.\ Journal of Cosmology and Astroparticle Physics 2026. doi:10.1088/1475-7516/2026/05/073

\bibitem[Cole et al.(2005)]{2005MNRAS.362..505C} Cole, S. and 30 colleagues 2005.\ The 2dF Galaxy Redshift Survey: power-spectrum analysis of the final data set and cosmological implications.\ Monthly Notices of the Royal Astronomical Society 362, 505--534. doi:10.1111/j.1365-2966.2005.09318.x

\bibitem[Coulton et al.(2024)]{2024PhRvD.109f3530C} Coulton, W. and 153 colleagues 2024.\ Atacama Cosmology Telescope: High-resolution component-separated maps across one third of the sky.\ Physical Review D 109. doi:10.1103/PhysRevD.109.063530

\bibitem[Das et al.(2011)]{2011PhRvL.107b1301D} Das, S. and 40 colleagues 2011.\ Detection of the Power Spectrum of Cosmic Microwave Background Lensing by the Atacama Cosmology Telescope.\ Physical Review Letters 107. doi:10.1103/PhysRevLett.107.021301
  
\bibitem[d'Assignies et al.(2025)]{2025arXiv251023565D} d'Assignies, W. and 94 colleagues 2025.\ Dark Energy Survey Year 6 Results: Clustering-redshifts and importance sampling of Self-Organised-Maps $n(z)$ realizations for $3\times2$pt samples.\ arXiv e-prints. doi:10.48550/arXiv.2510.23565

\bibitem[DES Collaboration et al.(2026)]{2026arXiv260210065D} DES Collaboration and 128 colleagues 2026.\ Dark Energy Survey Year 6 Results: Cosmological Constraints from Cosmic Shear.\ arXiv e-prints. doi:10.48550/arXiv.2602.10065

\bibitem[DES Collaboration et al.(2026)]{2026arXiv260114559D} DES Collaboration and 171 colleagues 2026.\ Dark Energy Survey Year 6 Results: Cosmological Constraints from Galaxy Clustering and Weak Lensing.\ arXiv e-prints. doi:10.48550/arXiv.2601.14559
  
\bibitem[Di Valentino et al.(2025)]{2025PDU....4901965D} Di Valentino, E. and 199 colleagues 2025.\ The CosmoVerse White Paper: Addressing observational tensions in cosmology with systematics and fundamental physics.\ Physics of the Dark Universe 49. doi:10.1016/j.dark.2025.101965

\bibitem[Ding et al.(2024)]{2024MNRAS.527.3728D} Ding, Z., Yu, Y., Zhang, P.\ 2024.\ Fisher forecast for the BAO measurements from the CSST spectroscopic and photometric galaxy clustering.\ Monthly Notices of the Royal Astronomical Society 527, 3728--3740. doi:10.1093/mnras/stad3379

\bibitem[Dong et al.(2022)]{2022ApJ...938...72D} Dong, F., Zhang, P., Sun, Z., Park, C.\ 2022.\ The First Direct Measurement of Gravitational Potential Decay Rate at Cosmological Scales and Improved Dark Energy Constraint.\ The Astrophysical Journal 938. doi:10.3847/1538-4357/ac905b

\bibitem[Efstathiou et al.(1990)]{1990Natur.348..705E} Efstathiou, G., Sutherland, W.~J., Maddox, S.~J.\ 1990.\ The cosmological constant and cold dark matter.\ Nature 348, 705--707. doi:10.1038/348705a0

\bibitem[Eisenstein et al.(2005)]{2005ApJ...633..560E} Eisenstein, D.~J. and 47 colleagues 2005.\ Detection of the Baryon Acoustic Peak in the Large-Scale Correlation Function of SDSS Luminous Red Galaxies.\ The Astrophysical Journal 633, 560--574. doi:10.1086/466512

\bibitem[Fan et al.(2010)]{2010ApJ...719.1408F} Fan, Z., Shan, H., Liu, J.\ 2010.\ Noisy Weak-lensing Convergence Peak Statistics Near Clusters of Galaxies and Beyond.\ The Astrophysical Journal 719, 1408--1420. doi:10.1088/0004-637X/719/2/1408

\bibitem[Feng et al.(2005)]{2005PhLB..607...35F} Feng, B., Wang, X., Zhang, X.\ 2005.\ Dark energy constraints from the cosmic age and supernova.\ Physics Letters B 607, 35--41. doi:10.1016/j.physletb.2004.12.071

\bibitem[Ferreira(2019)]{2019ARA&A..57..335F} Ferreira, P.~G.\ 2019.\ Cosmological Tests of Gravity.\ Annual Review of Astronomy and Astrophysics 57, 335--374. doi:10.1146/annurev-astro-091918-104423

\bibitem[Ge et al.(2025)]{2025PhRvD.111h3534G} Ge, F. and 96 colleagues 2025.\ Cosmology from CMB lensing and delensed EE power spectra using 2019--2020 SPT-3G polarization data.\ Physical Review D 111. doi:10.1103/PhysRevD.111.083534

\bibitem[Ghirardini et al.(2024)]{2024A&A...689A.298G} Ghirardini, V. and 48 colleagues 2024.\ The SRG/eROSITA all-sky survey: Cosmology constraints from cluster abundances in the western Galactic hemisphere.\ Astronomy and Astrophysics 689. doi:10.1051/0004-6361/202348852

\bibitem[Gu et al.(2025)]{2025NatAs...9.1879G} Gu, G. and 59 colleagues 2025.\ Dynamical dark energy in light of the DESI DR2 baryonic acoustic oscillations measurements.\ Nature Astronomy 9, 1879--1889. doi:10.1038/s41550-025-02669-6

\bibitem[Gubitosi et al.(2013)]{2013JCAP...02..032G} Gubitosi, G., Piazza, F., Vernizzi, F.\ 2013.\ The effective field theory of dark energy.\ Journal of Cosmology and Astroparticle Physics 2013. doi:10.1088/1475-7516/2013/02/032
  
\bibitem[Hadzhiyska et al.(2025)]{2025PhRvD.112h3509H} Hadzhiyska, B. and 75 colleagues 2025.\ Evidence for large baryonic feedback at low and intermediate redshifts from kinematic Sunyaev-Zel'dovich observations with ACT and DESI photometric galaxies.\ Physical Review D 112. doi:10.1103/kclp-x5j1
  
\bibitem[Hadzhiyska et al.(2026)]{2026arXiv260419745H} Hadzhiyska, B. and 52 colleagues 2026.\ Precision Kinematic Sunyaev--Zel'dovich Measurements Across Halo Mass and Redshift with DESI DR2 and ACT DR6: Part II. Bright Galaxy Survey and Emission-Line Galaxies.\ arXiv e-prints. doi:10.48550/arXiv.2604.19745

\bibitem[Hand et al.(2012)]{2012PhRvL.109d1101H} Hand, N. and 57 colleagues 2012.\ Evidence of Galaxy Cluster Motions with the Kinematic Sunyaev-Zel'dovich Effect.\ Physical Review Letters 109. doi:10.1103/PhysRevLett.109.041101

\bibitem[Ho et al.(2009)]{2009arXiv0903.2845H} Ho, S., Dedeo, S., Spergel, D.\ 2009.\ Finding the Missing Baryons Using CMB as a Backlight.\ arXiv e-prints. doi:10.48550/arXiv.0903.2845

\bibitem[Hu et al.(2014)]{2014PhRvD..89j3530H} Hu, B., Raveri, M., Frusciante, N., Silvestri, A.\ 2014.\ Effective field theory of cosmic acceleration: An implementation in CAMB.\ Physical Review D 89. doi:10.1103/PhysRevD.89.103530

\bibitem[Ishak(2019)]{2019LRR....22....1I} Ishak, M.\ 2019.\ Testing general relativity in cosmology.\ Living Reviews in Relativity 22. doi:10.1007/s41114-018-0017-4

\bibitem[Ishak et al.(2025)]{2025JCAP...09..053I} Ishak, M. and 70 colleagues 2025.\ Modified gravity constraints from the full shape modeling of clustering measurements from DESI 2024.\ Journal of Cosmology and Astroparticle Physics 2025. doi:10.1088/1475-7516/2025/09/053

\bibitem[Jain and Taylor(2003)]{2003PhRvL..91n1302J} Jain, B., Taylor, A.\ 2003.\ Cross-Correlation Tomography: Measuring Dark Energy Evolution with Weak Lensing.\ Physical Review Letters 91. doi:10.1103/PhysRevLett.91.141302

\bibitem[Jain and Zhang(2008)]{2008PhRvD..78f3503J} Jain, B., Zhang, P.\ 2008.\ Observational tests of modified gravity.\ Physical Review D 78. doi:10.1103/PhysRevD.78.063503

\bibitem[Jeong et al.(2012)]{2012PhRvD..85b3504J} Jeong, D., Schmidt, F., Hirata, C.~M.\ 2012.\ Large-scale clustering of galaxies in general relativity.\ Physical Review D 85. doi:10.1103/PhysRevD.85.023504

\bibitem[Joyce et al.(2015)]{2015PhR...568....1J} Joyce, A., Jain, B., Khoury, J., Trodden, M.\ 2015.\ Beyond the cosmological standard model.\ Physics Reports 568, 1--98. doi:10.1016/j.physrep.2014.12.002

\bibitem[Joyce et al.(2016)]{2016ARNPS..66...95J} Joyce, A., Lombriser, L., Schmidt, F.\ 2016.\ Dark Energy Versus Modified Gravity.\ Annual Review of Nuclear and Particle Science 66, 95--122. doi:10.1146/annurev-nucl-102115-044553

\bibitem[Klein et al.(2019)]{2019MNRAS.488..739K} Klein, M. and 65 colleagues 2019.\ A new RASS galaxy cluster catalogue with low contamination extending to z {\ensuremath{\sim}} 1 in the DES overlap region.\ Monthly Notices of the Royal Astronomical Society 488, 739--769. doi:10.1093/mnras/stz1463

\bibitem[Krauss and Turner(1995)]{1995GReGr..27.1137K} Krauss, L.~M., Turner, M.~S.\ 1995.\ The cosmological constant is back.\ General Relativity and Gravitation 27, 1137--1144. doi:10.1007/BF02108229
  
\bibitem[Krywonos et al.(2024)]{2024arXiv240805264K} Krywonos, J., Hotinli, S.~C., Johnson, M.~C.\ 2024.\ Constraints on cosmology beyond $\Lambda$CDM with kinetic Sunyaev Zel'dovich velocity reconstruction.\ arXiv e-prints. doi:10.48550/arXiv.2408.05264

\bibitem[Kunz and Sapone(2007)]{2007PhRvL..98l1301K} Kunz, M., Sapone, D.\ 2007.\ Dark Energy versus Modified Gravity.\ Physical Review Letters 98. doi:10.1103/PhysRevLett.98.121301
  
\bibitem[La Posta et al.(2025)]{2025PhRvD.112d3525L} La Posta, A., Alonso, D., Chisari, N.~E., Ferreira, T., Garc{\'\i}a-Garc{\'\i}a, C.\ 2025.\ Insights on gas thermodynamics from the combination of x-ray and thermal Sunyaev-Zel'dovich data cross correlated with cosmic shear.\ Physical Review D 112. doi:10.1103/m77z-w7pl
  
\bibitem[Li et al.(2026)]{2026JCAP...02..065L} Li, J., Zheng, Y., Zhu, W.\ 2026.\ Tracing missing baryons in the cosmic filaments with tSZ and CMB-lensing stacking.\ Journal of Cosmology and Astroparticle Physics 2026. doi:10.1088/1475-7516/2026/02/065

\bibitem[Li et al.(2023)]{2023PhRvD.108l3518L} Li, X. and 34 colleagues 2023.\ Hyper Suprime-Cam Year 3 results: Cosmology from cosmic shear two-point correlation functions.\ Physical Review D 108. doi:10.1103/PhysRevD.108.123518

\bibitem[Libanore et al.(2022)]{2022JCAP...02..003L} Libanore, S. and 7 colleagues 2022.\ Clustering of Gravitational Wave and Supernovae events: a multitracer analysis in Luminosity Distance Space.\ Journal of Cosmology and Astroparticle Physics 2022. doi:10.1088/1475-7516/2022/02/003

\bibitem[Linder(2003)]{2003PhRvL..90i1301L} Linder, E.~V.\ 2003.\ Exploring the Expansion History of the Universe.\ Physical Review Letters 90. doi:10.1103/PhysRevLett.90.091301

\bibitem[Linder(2005)]{2005PhRvD..72d3529L} Linder, E.~V.\ 2005.\ Cosmic growth history and expansion history.\ Physical Review D 72. doi:10.1103/PhysRevD.72.043529

\bibitem[Liu et al.(2025)]{2025PhRvD.112h3561L} Liu, R.~H. and 56 colleagues 2025.\ Measurements of the thermal Sunyaev-Zel'dovich effect with ACT and DESI luminous red galaxies.\ Physical Review D 112. doi:10.1103/jqn8-19gx

\bibitem[Lodha et al.(2025)]{2025PhRvD.112h3511L} Lodha, K. and 126 colleagues 2025.\ Extended dark energy analysis using DESI DR2 BAO measurements.\ Physical Review D 112. doi:10.1103/w4c6-1r5j

\bibitem[Mainieri et al.(2024)]{2024arXiv240305398M} Mainieri, V. and 199 colleagues 2024.\ The Wide-field Spectroscopic Telescope (WST) Science White Paper.\ arXiv e-prints. doi:10.48550/arXiv.2403.05398

\bibitem[Maus et al.(2025)]{2025JCAP...11..077M} Maus, M. and 60 colleagues 2025.\ A joint analysis of 3D clustering and galaxy {\texttimes} CMB-lensing cross-correlations with DESI DR1 galaxies.\ Journal of Cosmology and Astroparticle Physics 2025. doi:10.1088/1475-7516/2025/11/077

\bibitem[McQuinn(2014)]{2014ApJ...780L..33M} McQuinn, M.\ 2014.\ Locating the ``Missing'' Baryons with Extragalactic Dispersion Measure Estimates.\ The Astrophysical Journal 780. doi:10.1088/2041-8205/780/2/L33
  
\bibitem[Newman(2008)]{2008ApJ...684...88N} Newman, J.~A.\ 2008.\ Calibrating Redshift Distributions beyond Spectroscopic Limits with Cross-Correlations.\ The Astrophysical Journal 684, 88--101. doi:10.1086/589982

\bibitem[Oguri and Miyazaki(2025)]{2025PJAB..101..129O} Oguri, M., Miyazaki, S.\ 2025.\ Peaks in weak lensing mass maps for cluster astrophysics and cosmology.\ Proceedings of the Japan Academy, Series B 101, 129--142. doi:10.2183/pjab.101.010

\bibitem[Peebles(1980)]{1980lssu.book.....P} Peebles, P.~J.~E.\ 1980.\ The large-scale structure of the universe.\ Large-Scale Structure of the Universe by Phillip James Edwin Peebles. Princeton University Press, 1980. ISBN: 978-0-691-08240-0.

\bibitem[Peebles(1984)]{1984ApJ...284..439P} Peebles, P.~J.~E.\ 1984.\ Tests of cosmological models constrained by inflation.\ The Astrophysical Journal 284, 439--444. doi:10.1086/162425

\bibitem[Perlmutter et al.(1999)]{1999ApJ...517..565P} Perlmutter, S. and 32 colleagues 1999.\ Measurements of {\ensuremath{\Omega}} and {\ensuremath{\Lambda}} from 42 High-Redshift Supernovae.\ The Astrophysical Journal 517, 565--586. doi:10.1086/307221

\bibitem[Planck Collaboration et al.(2014)]{2014A&A...571A..16P} Planck Collaboration and 199 colleagues 2014.\ Planck 2013 results. XVI. Cosmological parameters.\ Astronomy and Astrophysics 571. doi:10.1051/0004-6361/201321591
  
\bibitem[Planck Collaboration et al.(2014)]{2014A&A...561A..97P} Planck Collaboration and 184 colleagues 2014.\ Planck intermediate results. XIII. Constraints on peculiar velocities.\ Astronomy and Astrophysics 561. doi:10.1051/0004-6361/201321299

\bibitem[Planck Collaboration et al.(2016)]{2016A&A...594A..22P} Planck Collaboration and 199 colleagues 2016.\ Planck 2015 results. XXII. A map of the thermal Sunyaev-Zeldovich effect.\ Astronomy and Astrophysics 594. doi:10.1051/0004-6361/201525826
  
\bibitem[Planck Collaboration et al.(2016a)]{2016A&A...594A..27P} Planck Collaboration and 199 colleagues 2016.\ Planck 2015 results. XXVII. The second Planck catalogue of Sunyaev-Zeldovich sources.\ Astronomy and Astrophysics 594. doi:10.1051/0004-6361/201525823

\bibitem[Planck Collaboration et al.(2016b)]{Planck2015cluster} Planck Collaboration and 199 colleagues 2016.\ Planck 2015 results. XXIV. Cosmology from Sunyaev-Zeldovich cluster counts.\ Astronomy and Astrophysics 594. doi:10.1051/0004-6361/201525833

\bibitem[Planck Collaboration et al.(2020)]{2020A&A...641A...1P} Planck Collaboration and 194 colleagues 2020.\ Planck 2018 results. I. Overview and the cosmological legacy of Planck.\ Astronomy and Astrophysics 641. doi:10.1051/0004-6361/201833880

\bibitem[Qin et al.(2025)]{2025PhRvD.112d3522Q} Qin, J. and 6 colleagues 2025.\ Weak lensing reconstruction by counting galaxies: Improvement with DES Y3 galaxies.\ Physical Review D 112. doi:10.1103/bwyy-xq85

\bibitem[Qin et al.(2026)]{2026ApJ..1001..121Q} Qin, J. and 6 colleagues 2026.\ Exploring the S$_{8}$ Tension: Insights from the CatNorth 1.5 Million Quasar Candidates.\ The Astrophysical Journal 1001. doi:10.3847/1538-4357/ae4ee5

\bibitem[Qin et al.(2026)]{Qin26} Qin, F. and 65 colleagues 2026.\ The DESI DR1 peculiar velocity survey: Growth rate measurements from the galaxy power spectrum.\ Astronomy and Astrophysics 708. doi:10.1051/0004-6361/202558368

\bibitem[Qu et al.(2024)]{2024ApJ...962..112Q} Qu, F.~J. and 157 colleagues 2024.\ The Atacama Cosmology Telescope: A Measurement of the DR6 CMB Lensing Power Spectrum and Its Implications for Structure Growth.\ The Astrophysical Journal 962. doi:10.3847/1538-4357/acfe06

\bibitem[Qu et al.(2026)]{2026PhRvL.136b1001Q} Qu, F.~J. and 146 colleagues 2026.\ Unified and Consistent Structure Growth Measurements from Joint ACT, SPT, and Planck CMB Lensing.\ Physical Review Letters 136. doi:10.1103/k5yr-3h6d

\bibitem[Qu et al.(2026)]{2026arXiv260419744Q} Qu, F.~J. and 53 colleagues 2026.\ Precision Kinematic Sunyaev--Zel'dovich Measurements Across Halo Mass and Redshift with DESI DR2 and ACT DR6: Part I. Luminous Red Galaxies.\ arXiv e-prints. doi:10.48550/arXiv.2604.19744

\bibitem[Raghunathan et al.(2026)]{2026ApJ..1002L..22R} Raghunathan, S. and 143 colleagues 2026.\ Measurement of the Full Shape of the Thermal Sunyaevâ--Zel'dovich Power Spectrum from the South Pole Telescope and Herschel--SPIRE Observations.\ The Astrophysical Journal 1002. doi:10.3847/2041-8213/ae5c07

\bibitem[Riess et al.(1998)]{1998AJ....116.1009R} Riess, A.~G. and 19 colleagues 1998.\ Observational Evidence from Supernovae for an Accelerating Universe and a Cosmological Constant.\ The Astronomical Journal 116, 1009--1038. doi:10.1086/300499

\bibitem[Sailer et al.(2025)]{2025PhRvD.111j3540S} Sailer, N. and 7 colleagues 2025.\ Evolution of structure growth during dark energy domination: Insights from the cross-correlation of DESI galaxies with CMB lensing and galaxy magnification.\ Physical Review D 111. doi:10.1103/27rg-tq8z

\bibitem[S{\'a}nchez et al.(2022)]{2022PhRvD.105h3529S} S{\'a}nchez, C. and 141 colleagues 2022.\ Dark Energy Survey Year 3 results: Exploiting small-scale information with lensing shear ratios.\ Physical Review D 105. doi:10.1103/PhysRevD.105.083529
  
\bibitem[Schneider et al.(2006)]{2006ApJ...651...14S} Schneider, M., Knox, L., Zhan, H., Connolly, A.\ 2006.\ Using Galaxy Two-Point Correlation Functions to Determine the Redshift Distributions of Galaxies Binned by Photometric Redshift.\ The Astrophysical Journal 651, 14--23. doi:10.1086/507675

\bibitem[Scolnic et al.(2018)]{2018ApJ...859..101S} Scolnic, D.~M. and 39 colleagues 2018.\ The Complete Light-curve Sample of Spectroscopically Confirmed SNe Ia from Pan-STARRS1 and Cosmological Constraints from the Combined Pantheon Sample.\ The Astrophysical Journal 859. doi:10.3847/1538-4357/aab9bb

\bibitem[Scolnic et al.(2022)]{2022ApJ...938..113S} Scolnic, D. and 29 colleagues 2022.\ The Pantheon+ Analysis: The Full Data Set and Light-curve Release.\ The Astrophysical Journal 938. doi:10.3847/1538-4357/ac8b7a

\bibitem[Scranton et al.(2005)]{2005ApJ...633..589S} Scranton, R. and 17 colleagues 2005.\ Detection of Cosmic Magnification with the Sloan Digital Sky Survey.\ The Astrophysical Journal 633, 589--602. doi:10.1086/431358
  
\bibitem[Shao et al.(2011)]{2011MNRAS.413..628S} Shao, J., Zhang, P., Lin, W., Jing, Y., Pan, J.\ 2011.\ Kinetic Sunyaev-Zel'dovich tomography with spectroscopic redshift surveys.\ Monthly Notices of the Royal Astronomical Society 413, 628--642. doi:10.1111/j.1365-2966.2011.18166.x

\bibitem[Sherwin et al.(2011)]{2011PhRvL.107b1302S} Sherwin, B.~D. and 37 colleagues 2011.\ Evidence for Dark Energy from the Cosmic Microwave Background Alone Using the Atacama Cosmology Telescope Lensing Measurements.\ Physical Review Letters 107. doi:10.1103/PhysRevLett.107.021302

\bibitem[Shi et al.(2024)]{2024MNRAS.528.4922S} Shi, Y., Zhang, P., Mao, S., Gu, Q.\ 2024.\ Momentum power spectrum of SDSS galaxies by massE cosmic ruler: 2.1{\texttimes}improvement in measure of growth rate.\ Monthly Notices of the Royal Astronomical Society 528, 4922--4940. doi:10.1093/mnras/stae274

\bibitem[Okumura and Taruya(2022)]{2022PhRvD.106d3523O} Okumura, T., Taruya, A.\ 2022.\ Tightening geometric and dynamical constraints on dark energy and gravity: Galaxy clustering, intrinsic alignment, and kinetic Sunyaev-Zel'dovich effect.\ Physical Review D 106. doi:10.1103/PhysRevD.106.043523
  
\bibitem[Peng and Yu(2024)]{2024JCAP...10..025P} Peng, H., Yu, Y.\ 2024.\ Reconstructing redshift distributions with photometric galaxy clustering.\ Journal of Cosmology and Astroparticle Physics 2024. doi:10.1088/1475-7516/2024/10/025

\bibitem[Tully et al.(2023)]{2023ApJ...944...94T} Tully, R.~B. and 16 colleagues 2023.\ Cosmicflows-4.\ The Astrophysical Journal 944. doi:10.3847/1538-4357/ac94d8

\bibitem[Weinberg et al.(2013)]{2013PhR...530...87W} Weinberg, D.~H., Mortonson, M.~J., Eisenstein, D.~J., Hirata, C., Riess, A.~G., Rozo, E.\ 2013.\ Observational probes of cosmic acceleration.\ Physics Reports 530, 87--255. doi:10.1016/j.physrep.2013.05.001

\bibitem[Wojtak et al.(2011)]{2011Natur.477..567W} Wojtak, R., Hansen, S.~H., Hjorth, J.\ 2011.\ Gravitational redshift of galaxies in clusters as predicted by general relativity.\ Nature 477, 567--569. doi:10.1038/nature10445

 \bibitem[Xu et al.(2023)]{2023NatAs...7.1259X} Xu, K., Jing, Y.~P., Zhao, G.-B., Cuesta, A.~J.\ 2023.\ Evidence for baryon acoustic oscillations from galaxy-ellipticity correlations..\ Nature Astronomy 7, 1259--1264. doi:10.1038/s41550-023-02035-4

\bibitem[Xu et al.(2024)]{2024ApJ...973..102X} Xu, K., Jing, Y.~P., Gao, H., Luo, X., Li, M.\ 2024.\ Accurate Measurement of the Lensing Magnification by BOSS CMASS Galaxies and Its Implications for Cosmology and Dark Matter.\ The Astrophysical Journal 973. doi:10.3847/1538-4357/ad6156

\bibitem[Yang et al.(2021)]{2021ApJ...909..143Y} Yang, X. and 22 colleagues 2021.\ An Extended Halo-based Group/Cluster Finder: Application to the DESI Legacy Imaging Surveys DR8.\ The Astrophysical Journal 909. doi:10.3847/1538-4357/abddb2

\bibitem[Yao et al.(2024)]{2024MNRAS.527.5206Y} Yao, J. and 13 colleagues 2024.\ CSST WL preparation I: forecast the impact from non-Gaussian covariances and requirements on systematics control.\ Monthly Notices of the Royal Astronomical Society 527, 5206--5218. doi:10.1093/mnras/stad3563

\bibitem[Zaborowski et al.(2026)]{2026JCAP...04..004Z} Zaborowski, E.~A. and 60 colleagues 2026.\ H $_{0}$ without the sound horizon (or supernovae): A 2\% measurement in DESI DR1.\ Journal of Cosmology and Astroparticle Physics 2026. doi:10.1088/1475-7516/2026/04/004
  
\bibitem[Zhang et al.(2005)]{2005ApJ...635..806Z} Zhang, J., Hui, L., Stebbins, A.\ 2005.\ Isolating Geometry in Weak-Lensing Measurements.\ The Astrophysical Journal 635, 806--820. doi:10.1086/497676
  
\bibitem[Zhang et al.(2017)]{2017ApJ...848...44Z} Zhang, L., Yu, Y., Zhang, P.\ 2017.\ Non-negative Matrix Factorization for Self-calibration of Photometric Redshift Scatter in Weak-lensing Surveys.\ The Astrophysical Journal 848. doi:10.3847/1538-4357/aa8c72

\bibitem[Zhang and Pen(2001)]{2001ApJ...549...18Z} Zhang, P., Pen, U.-L.\ 2001.\ Deprojecting Sunyaev-Zeldovich Statistics.\ The Astrophysical Journal 549, 18--27. doi:10.1086/319067

\bibitem[Zhang et al.(2007)]{2007PhRvL..99n1302Z} Zhang, P., Liguori, M., Bean, R., Dodelson, S.\ 2007.\ Probing Gravity at Cosmological Scales by Measurements which Test the Relationship between Gravitational Lensing and Matter Overdensity.\ Physical Review Letters 99. doi:10.1103/PhysRevLett.99.141302
  
\bibitem[Zhang et al.(2010)]{2010MNRAS.405..359Z} Zhang, P., Pen, U.-L., Bernstein, G.\ 2010.\ Self-calibration of photometric redshift scatter in weak-lensing surveys.\ Monthly Notices of the Royal Astronomical Society 405, 359--374. doi:10.1111/j.1365-2966.2010.16445.x

\bibitem[Zhang and Stebbins(2011)]{2011PhRvL.107d1301Z} Zhang, P., Stebbins, A.\ 2011.\ Confirmation of the Copernican Principle at Gpc Radial Scale and above from the Kinetic Sunyaev-Zel'dovich Effect Power Spectrum.\ Physical Review Letters 107. doi:10.1103/PhysRevLett.107.041301
  
\bibitem[Zhang and Johnson(2015)]{2015JCAP...06..046Z} Zhang, P., Johnson, M.~C.\ 2015.\ Testing eternal inflation with the kinetic Sunyaev Zel'dovich effect.\ Journal of Cosmology and Astroparticle Physics 2015, 046--046. doi:10.1088/1475-7516/2015/06/046

\bibitem[Zhang(2018)]{2018arXiv181011915Z} Zhang, P.\ 2018.\ The large scale structure in the 3D luminosity-distance space and its cosmological applications.\ arXiv e-prints. doi:10.48550/arXiv.1810.11915

\bibitem[Zhao et al.(2021)]{2021MNRAS.504...33Z} Zhao, G.-B. and 25 colleagues 2021.\ The completed SDSS-IV extended Baryon Oscillation Spectroscopic Survey: a multitracer analysis in Fourier space for measuring the cosmic structure growth and expansion rate.\ Monthly Notices of the Royal Astronomical Society 504, 33--52. doi:10.1093/mnras/stab849

\bibitem[Zhao et al.(2022)]{2022MNRAS.511.5492Z} Zhao, C. and 17 colleagues 2022.\ The completed SDSS-IV extended Baryon Oscillation Spectroscopic Survey: cosmological implications from multitracer BAO analysis with galaxies and voids.\ Monthly Notices of the Royal Astronomical Society 511, 5492--5524. doi:10.1093/mnras/stac390

\bibitem[Zhao et al.(2024)]{2024arXiv241107970Z} Zhao, C. and 43 colleagues 2024.\ MUltiplexed Survey Telescope: Perspectives for Large-Scale Structure Cosmology in the Era of Stage-V Spectroscopic Survey.\ arXiv e-prints. doi:10.48550/arXiv.2411.07970

\bibitem[Zhou and Zhang(2025)]{2025arXiv251011022Z} Zhou, S., Zhang, P.\ 2025.\ Cavendish experiment with fast radio bursts on cosmological scales.\ arXiv e-prints. doi:10.48550/arXiv.2510.11022

\end{thebibliography}
\end{document}